\documentclass[10pt]{article}

\usepackage[a4paper,margin=2.15cm]{geometry}
\usepackage{amsmath,amssymb}
\usepackage{booktabs}
\usepackage{graphicx}
\usepackage{microtype}
\usepackage{xcolor}
\usepackage{hyperref}
\usepackage{authblk}
\usepackage{caption}
\usepackage{enumitem}
\usepackage{array}
\usepackage{tabularx}

\definecolor{linkblue}{HTML}{175F82}
\hypersetup{colorlinks=true,allcolors=linkblue,pdfauthor={Jose Antonio Sanchez-Andreu}}
\setlist{nosep,leftmargin=*}
\title{Auditing Representation-Induced Geometry in\\Acoustic-Emission Fracture Transients}
\author[1]{Jose Antonio Sanchez-Andreu}
\affil[1]{Independent Researcher, Spain\\ORCID: \href{https://orcid.org/0009-0006-0306-7253}{0009-0006-0306-7253}}
\date{September 2026}

\begin{document}
\maketitle

\begin{abstract}
Fixed convolutional representations can induce structured geometry in time--frequency data even without task-specific or source-domain training. We audit a previously reported morphology-space analysis of two granite acoustic-emission experiments. The original cumulative angular-path endpoint is reproduced to within \(2.1\times10^{-7}\) in the Archive-B/Archive-A ratio. However, the same directional contrast is obtained with ImageNet features and with three randomly initialized EfficientNet-B0 encoders: the preserved Aether representation gives a ratio of 1.411, ImageNet gives 1.361, and the random encoders give 1.480--1.503. We therefore withdraw the attribution of this contrast to interferometric pretraining. We also correct a reversal between archived OG1/OG3 file identifiers and the physical experiments. Phase randomization and temporal shuffling alter the preserved trajectory, but these responses are reinterpreted as pipeline-dependent diagnostics rather than evidence of a learned universal elastic morphology. The corrected result is narrower: multiple fixed convolutional maps expose a reproducible contrast between two acoustic-emission datasets. The signal statistics and architectural biases responsible for that contrast remain open questions.
\end{abstract}

\section*{Revision statement}

This version is a substantive correction and refocusing of version 1 of this arXiv record. Version 1 attributed a latent angular-path contrast between two granite acoustic-emission archives to a representation pretrained on Gravity Spy interferometric transients, and interpreted the contrast in terms of two named fracture regimes. A subsequent audit established two facts that materially change that interpretation:

\begin{enumerate}
  \item the numerical endpoint is reproducible with the preserved representation and preprocessing; and
  \item its direction is not specific to that representation: ImageNet and three random encoders reproduce it, while the archived file identifiers used in the analysis were reversed relative to the physical experiment identities.
\end{enumerate}

\begin{figure}[t]
  \centering
  \includegraphics[width=\textwidth]{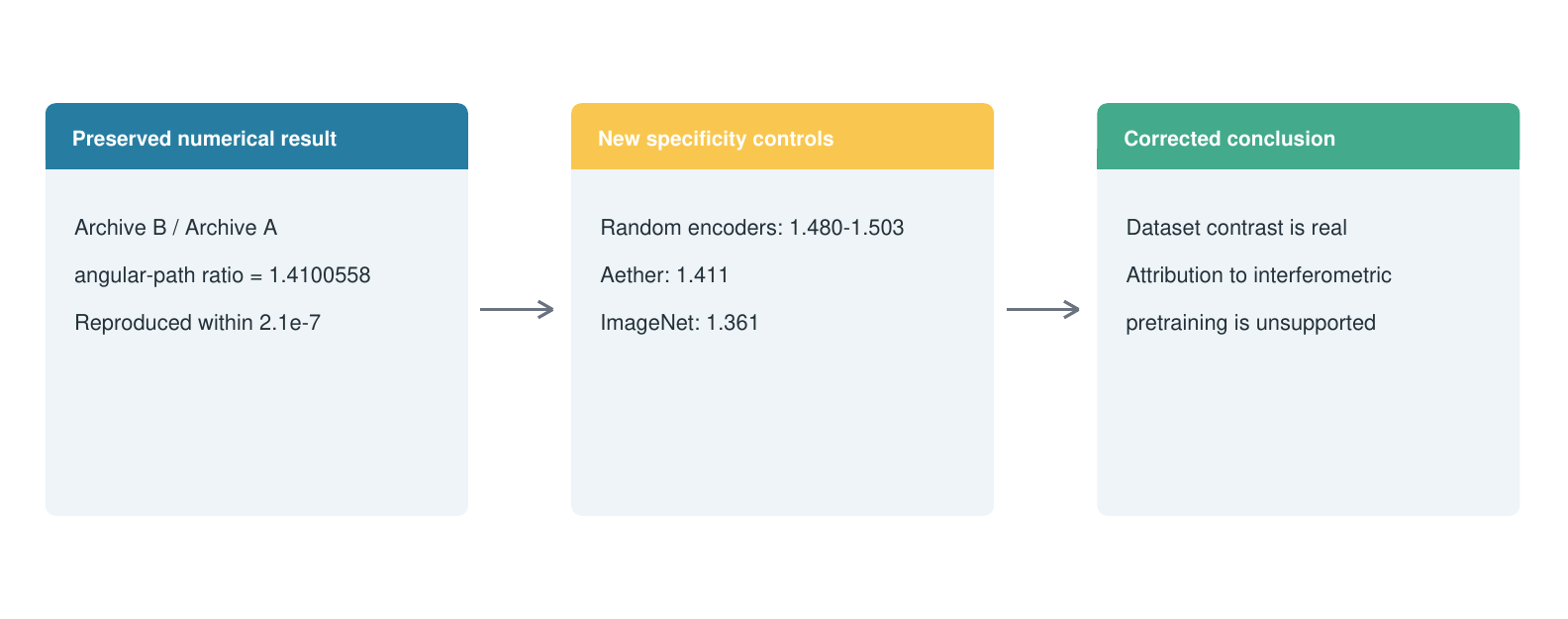}
  \caption{Outcome of the audit. The central numerical contrast survives exact reproduction, but encoder controls change its interpretation. The result is evidence for a dataset contrast under several fixed convolutional maps, not evidence that interferometric pretraining learned universal fracture physics.}
  \label{fig:auditlogic}
\end{figure}

Accordingly, this version withdraws the claims of learned cross-domain physical transfer, encoder specificity, and regime-specific causal interpretation. Analyses from version 1 that do not contain an encoder-matched control---including joint PCA, seismic classification, synthetic attribution maps, and cross-domain examples---are not used as evidence here. Figure~\ref{fig:auditlogic} summarizes the corrected logic.

\section{Background and corrected question}

Acoustic-emission waveforms record transient elastic energy release during rock fracture, but the mapping from waveform statistics to crack organization is experiment- and acquisition-dependent \cite{lockner1993role,lei2003how}. Version 1 asked whether a frozen convolutional encoder trained on Gravity Spy glitches \cite{zevin2017gravityspy} could serve as a cross-domain morphology probe. Each event was converted to a log-amplitude short-time Fourier transform (STFT), embedded by an EfficientNet-B0 feature extractor, \(L_2\)-normalized, and ordered by normalized experiment time.

For consecutive normalized embeddings \(\hat z_i\), the endpoint was the cumulative angular path
\begin{equation}
  \Theta=\sum_{i=2}^{n}\arccos\!\left[\operatorname{clip}
  \left(\hat z_i^{\mathsf T}\hat z_{i-1},-1,1\right)\right].
  \label{eq:theta}
\end{equation}
This is a property of the complete pipeline---preprocessing, architecture, weights, event selection, ordering, and normalization---rather than a direct mechanical observable. A larger \(\Theta\) means only that successive inputs trace a longer path in that representation.

The corrected scientific question is therefore not whether Aether discovered a universal elastic morphology. It is: \emph{does the observed two-archive contrast require the source-domain training, or does it arise more broadly under fixed convolutional representations?}

\section{Audit design}

\subsection{Endpoint reproduction}

The historical embeddings and endpoint were compared against a bridge implementation that reconstructs the same STFT-to-embedding path. We recorded the maximum and mean absolute embedding errors, the minimum cosine similarity, and the absolute difference in the final Archive-B/Archive-A ratio. This test isolates numerical reproducibility; it does not establish physical validity.

\subsection{Encoder-specificity control}

The same archive samples, ordering, preprocessing, \(L_2\) normalization, and angular-path equation were evaluated with five fixed representations: the preserved Aether encoder, an ImageNet-pretrained EfficientNet-B0, and three randomly initialized EfficientNet-B0 models with seeds 42, 2026, and 137. No target-domain training was performed. The primary control asks whether the direction \(\Theta_B>\Theta_A\) disappears without interferometric pretraining.

This is a necessary but limited control. Three random seeds can falsify a claim of specificity when all reproduce the claimed direction, but they are not sufficient to estimate a general random-feature distribution.

\subsection{Archive identity audit}

The file names used by the pipeline were cross-checked against the physical experiment metadata described by Lei \cite{lei2003how}. To avoid propagating the original ambiguity, we use neutral analysis labels:
\begin{itemize}
  \item \textbf{Archive A}: \texttt{OG1\_070130.zip}, corresponding to physical OG3 (dry; 61.4 MPa confinement);
  \item \textbf{Archive B}: \texttt{OG3\_070213.zip}, corresponding to physical OG1 (uniaxial; 0 MPa confinement).
\end{itemize}
The numerical ratio reported in version 1 was Archive B divided by Archive A. The file reversal leaves the number unchanged but reverses its physical assignment.

\subsection{Perturbation diagnostics}

A preserved 2,000-event analysis of Archive B compared the original trajectory with amplitude normalization, temporal reversal, Fourier-phase randomization, and temporal shuffling. These diagnostics test whether the endpoint changes when signal structure is altered. Because matched perturbation experiments were not run for the ImageNet and random encoders, they cannot identify a feature learned specifically by Aether.

\section{Results}

\subsection{The historical endpoint is numerically reproducible}

The historical analysis gave \(\Theta_A=12.44575364\) rad and \(\Theta_B=17.54920681\) rad, for \(\Theta_B/\Theta_A=1.4100557759\). The bridge reproduction gave 12.44575524 rad and 17.54920647 rad, for a ratio of 1.4100555675 (Table~\ref{tab:reproduction}). The absolute ratio difference was \(2.0843\times10^{-7}\). Across reproduced embeddings, the maximum absolute error was \(1.2636\times10^{-5}\), the mean absolute error was \(1.4796\times10^{-7}\), and the minimum cosine similarity was 0.9999997616.

\begin{table}[h]
\centering
\caption{Reproduction of the historical cumulative angular-path endpoint. Archive labels are used because the physical identities were reversed in the original file naming.}
\label{tab:reproduction}
\begin{tabular}{lrrr}
\toprule
Pipeline & Archive A (rad) & Archive B (rad) & B/A ratio \\
\midrule
Historical & 12.44575364 & 17.54920681 & 1.4100557759 \\
Bridge reproduction & 12.44575524 & 17.54920647 & 1.4100555675 \\
\bottomrule
\end{tabular}
\end{table}

Thus the endpoint did not disappear because of a checkpoint-reading or preprocessing mismatch in this analysis. The failure of the original claim occurs at the level of attribution, not numerical existence.

\subsection{The direction is reproduced without Aether training}

\begin{figure}[t]
  \centering
  \includegraphics[width=\textwidth]{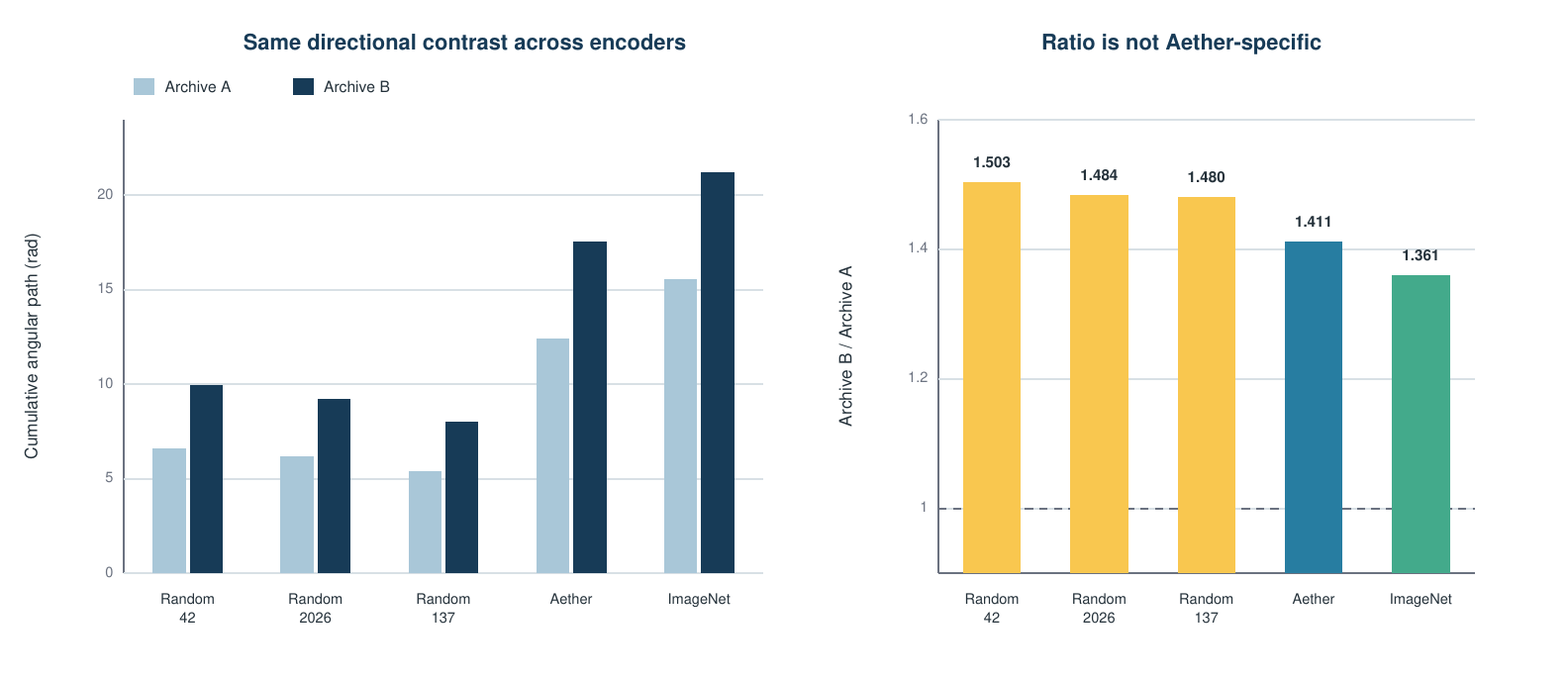}
  \caption{Encoder-specificity control. Left: cumulative angular paths differ in scale across representations. Right: all five encoders reproduce the Archive-B/Archive-A direction; the three random encoders yield larger ratios than Aether. This falsifies the claim that the direction requires interferometric pretraining.}
  \label{fig:specificity}
\end{figure}

Every controlled encoder produced \(\Theta_B>\Theta_A\) (Table~\ref{tab:specificity}; Figure~\ref{fig:specificity}). The three random encoders produced B/A ratios between 1.480 and 1.503, all larger than the Aether ratio of 1.411. ImageNet features produced a ratio of 1.361. Absolute path lengths differed substantially between encoders, as expected because each representation induces its own geometry, but the within-encoder direction was stable.

\begin{table}[h]
\centering
\caption{Encoder-specificity audit under the same endpoint. No target-domain training was used.}
\label{tab:specificity}
\begin{tabular}{lrrr}
\toprule
Representation & Archive A (rad) & Archive B (rad) & B/A ratio \\
\midrule
Random, seed 42 & 6.607469 & 9.933122 & 1.503317 \\
Random, seed 2026 & 6.208989 & 9.213476 & 1.483893 \\
Random, seed 137 & 5.415823 & 8.017699 & 1.480421 \\
Aether & 12.439771 & 17.555004 & 1.411200 \\
ImageNet & 15.579263 & 21.196027 & 1.360528 \\
\bottomrule
\end{tabular}
\end{table}

These controls support a narrower observation: the two archives differ in signal or acquisition statistics that multiple fixed convolutional maps expose. They do not identify which statistics are responsible, whether the contrast generalizes beyond these two experiments, or whether it tracks a fracture mechanism.

\subsection{The archive mapping changes the physical interpretation}

\begin{figure}[t]
  \centering
  \includegraphics[width=\textwidth]{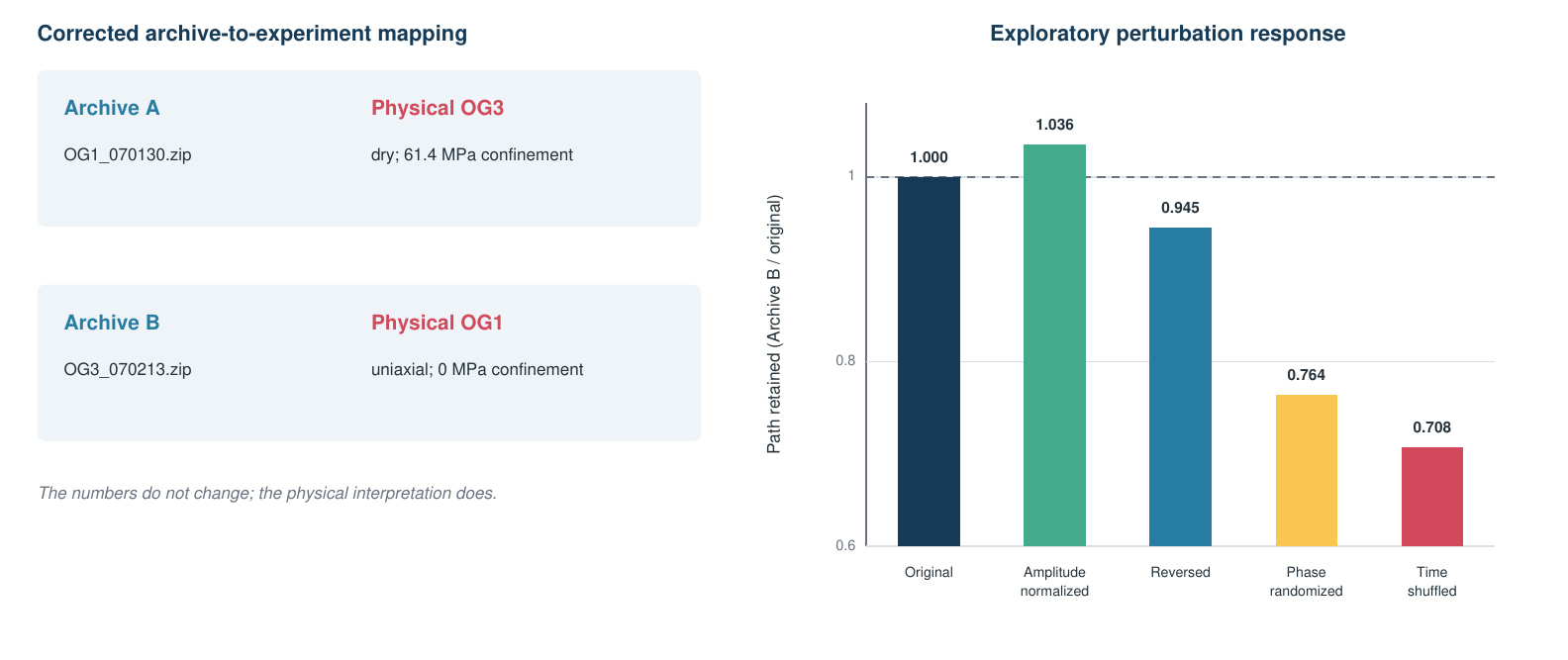}
  \caption{Left: corrected archive-to-experiment mapping. Right: preserved perturbation response for Archive B. Phase randomization and time shuffling shorten the path, but the result is exploratory because corresponding perturbation responses for ImageNet and random encoders were not measured.}
  \label{fig:mapping}
\end{figure}

Figure~\ref{fig:mapping} records the corrected mapping. In version 1, the archive name \texttt{OG3\_070213.zip} was interpreted as physical OG3; it is physical OG1. Conversely, \texttt{OG1\_070130.zip} is physical OG3. Consequently, statements in version 1 assigning the larger endpoint or perturbation sensitivity to a named localized or distributed regime are withdrawn. A physical interpretation would require reconstructing the experimental conditions from verified metadata and testing more than one trial per condition.

\subsection{Phase and time controls remain exploratory}

For the preserved 2,000-event Archive-B analysis, the original angular path was 1522.453 rad. Amplitude normalization gave 1576.628 rad (1.0356 of original), temporal reversal 1438.990 rad (0.9452), phase randomization 1163.488 rad (0.7642), and time shuffling 1077.345 rad (0.7076). These changes show that the particular pipeline is sensitive to the transformations, not that it learned a universal physical representation. The missing encoder-matched perturbation controls prevent a source-training attribution.

\section{What can and cannot be concluded}

\begin{table}[h]
\centering
\caption{Claim-level disposition after the audit.}
\label{tab:claims}
\begin{tabularx}{\textwidth}{>{\raggedright\arraybackslash}p{0.43\textwidth}>{\raggedright\arraybackslash}X}
\toprule
Supported by the preserved evidence & Withdrawn or unsupported \\
\midrule
The historical angular-path endpoint is reproduced with very high numerical agreement. & Interferometric/Gravity-Spy pretraining caused the two-archive contrast. \\
Multiple fixed convolutional maps yield a larger path for Archive B than Archive A. & The endpoint demonstrates a universal or shared physical morphology. \\
Phase and order perturbations change the preserved Archive-B trajectory. & Perturbation sensitivity is Aether-specific or identifies a fracture mechanism. \\
The archive labels were reversed relative to the physical experiment identities. & Version-1 assignments of the larger endpoint to a named physical regime. \\
\bottomrule
\end{tabularx}
\end{table}

The audit illustrates a general methodological point. A within-representation dataset contrast is not a test of representation provenance. To attribute an effect to source-domain pretraining, the endpoint must be compared against architecture-matched random weights, conventional pretraining, and direct signal baselines. Visualization controls are not substitutes for controls on the principal quantitative endpoint.

\section{Limitations and next tests}

This correction is intentionally narrow. First, only two acoustic-emission archives are compared; there are no replicate experiments from which to infer a population-level condition effect. Second, three random seeds falsify exclusivity but do not characterize the full random-encoder null distribution. Third, absolute angular-path scales are not directly comparable across latent spaces. Fourth, the perturbation analysis was not repeated across encoders. Fifth, data acquisition, event selection, and archive-specific preprocessing may confound the contrast. Finally, the Aether weights and derived artifacts are not yet public, so the present numerical audit cannot be independently repeated from this manuscript alone.

A prospective study should preregister an encoder-comparison endpoint, use at least 30 random seeds, include architecture-matched ImageNet and verified Aether weights, and compare against direct descriptors such as spectral moments, autocorrelation, permutation entropy, bicoherence, and wavelet/scattering coefficients. It should use verified physical metadata and multiple experiments per condition. Phase and order perturbations should be applied identically to every encoder and baseline. Such a study could determine whether the remaining contrast reflects interpretable signal statistics, generic convolutional bias, or fracture physics.

\section{Conclusion}

The main number in version 1 was not fabricated by a failed reproduction: the Archive-B/Archive-A angular-path ratio is recovered to within \(2.1\times10^{-7}\). What fails is the strong interpretation. ImageNet and three random EfficientNet-B0 encoders reproduce the direction, and the archive identifiers were reversed relative to the physical experiments. The defensible result is therefore a reproducible, representation-induced contrast between two acoustic-emission datasets. It is not evidence that interferometric pretraining learned universal elastic morphology. This version makes that distinction explicit and leaves the causal mechanism as a testable open question.

\section*{Data, code, and audit trail}

The numerical values reported here were recovered from preserved analysis summaries and reproduced embeddings associated with the audit. The original acoustic-emission source is described by Lei \cite{lei2003how}. A release containing the exact event manifests, model hashes, preprocessing code, per-encoder embeddings, and machine-readable audit tables is required for full external reproduction and will be linked to the arXiv record when available. Until that release exists, all claims in this version should be read as a transparent correction based on the preserved audit trail, not as an independently reproduced benchmark.

\section*{Acknowledgement of correction}

The author identified these issues during a post-publication internal reproducibility audit and takes responsibility for the overinterpretation and file-identity error in version 1. The purpose of this replacement is to preserve the valid numerical observation while correcting the scientific record.

\end{document}